\documentclass[usenatbib]{mn2e}
\usepackage{epsfig}
\usepackage{amssymb}

\begin{document}

\title[SCUBA map of the Groth Strip]{An $850\mathrm{\mu m}$ SCUBA map of the
Groth Strip and reliable source extraction}
\author[Coppin et al.]{
\parbox[t]{\textwidth}{
\vspace{-1.0cm}
Kristen Coppin$^{1}$, Mark Halpern $^{1}$, Douglas Scott$^{1}$, Colin Borys$^{2}$, Scott Chapman$^{2}$}
\vspace*{6pt}\\
$^{1}$ Department of Physics \& Astronomy, University of British Columbia, 
6224 Agricultural Road, Vancouver, BC, Canada  V6T 1Z1 \\
$^{2}$ California Institute of Technology, Pasadena, CA, USA  91125
\vspace*{-0.5cm}}

\date{Draft June 2004}

\maketitle

\begin{abstract}
We present an $850\,\mathrm{\mu m}$ map and list of candidate sources
in a sub-area of the Groth Strip observed using SCUBA.
The map consists of a long strip of adjoining
jiggle-maps covering the southwestern $70\,\mathrm{arcmin^{2}}$ of
the original WFPC2 Groth Strip to an average $1\,\sigma$ rms noise
level of $\simeq 3.5\,\mathrm{mJy}$.  We initially detect 7 candidate
sources with signal-to-noise ratios (SNRs) between 3.0 and
$3.5\,\sigma$ and 4 candidate sources with SNR $\geq3.5$.  Simulations
suggest that on average in a map this size one expects 1.6 false positive 
sources $\geq3.5\,\sigma$ and 4.5 between 3 and $3.5\,\sigma$.  Flux boosting 
in maps is a well known effect and we have developed a simple 
Bayesian prescription for estimating the unboosted flux distribution and used this 
method to determine the best flux estimates of our sources.  This method is easily 
adapted for any other modest signal-to-noise survey in which there is prior
knowledge of the source counts.  We performed 
follow-up photometry in an attempt to confirm or reject 5
of our source candidates.  We failed to significantly re-detect 3 of
the 5 sources in the noisiest regions of the map, suggesting that they
are either spurious or have true fluxes close to the noise level.
However, we did confirm the reality of 2 of the SCUBA sources, although at
lower flux levels than suggested in the map.  Not surprisingly, we find that the 
photometry results are consistent with and confirm the de-boosted map fluxes.
Our final candidate source list contains 3 sources, including the 2 confirmed detections and 1 further
candidate source with $\mathrm{SNR}>3.5\,\sigma$ which has a
reasonable chance of being real.  We performed correlations 
and found evidence of positive flux at the positions of 
{\sl XMM-Newton} X-ray sources.  The 95 per cent lower limit for 
the average flux density of these X-ray sources is $0.8\,\mathrm{mJy}$.
\end{abstract}

\begin{keywords}
submillimetre -- surveys -- cosmology: observations -- galaxies: high-redshift 
-- galaxies: starburst -- methods: statistical
\vspace*{-1.25cm}
\end{keywords}

\section{Introduction}
\label{sec:intro}
Large blank-field SCUBA surveys have revolutionized our
understanding of the importance and diverse nature of dusty galaxies
at high redshifts (e.g.~\citealt{Blain}, \citealt{Scott}, \citealt{Webb},
\citealt{Borys2003}).  Only about 300 blank-field SCUBA galaxies have been
discovered since the instrument was commissioned, in contrast to the
tens of thousands to millions of objects detected in optical surveys of similar
sizes.  Nevertheless, the number counts have been well-characterised,
and progress is being made in identifying SCUBA galaxies with objects
in other wavebands using source positions derived from radio identifications.

However, the fact remains that SCUBA sources are
difficult to find, and when they \textit{are} detected they typically
have low signal-to-noise ratios (SNRs), bringing into question the
reliability of measurements.  Using deep radio imaging of the 8-mJy Survey fields,
\citet{Ivison} have suggested that low SNR sources
in relatively noisy regions of submillimetre maps which lack radio
counterparts are often spurious.  Additional evidence that these sources
might be spurious comes from the lack of MAMBO (Max Planck Millimeter
Bolometer array) counterparts to many of these SCUBA sources
(\citealt{Greve}).  \citet{Mortier} also
report not recovering a similar fraction of low SNR sources in the
8-mJy Survey region when combined with newer SHADES (SCUBA HAlf
Degree Extragalactic Survey) data in the same field.
Only one region, namely the Hubble Deep Field (HDF) region or GOODS-North
field, has been investigated independently by different groups.  Reassuringly,
\citet{Borys2003, Borys2004} and \citet{Wang} are in close
agreement for the higher SNR sources in that region.  However, there is some
disagreement regarding the reality and the flux densities of several of the
noisier sources.  The effects of flux boosting (sometimes called
Malmquist and/or Eddington bias) are well known for SNR-thresholded maps,
and it is worthwhile to investigate whether such discrepancies are to be
expected, and when one should be confident about the reality of a source,
\textit{independent} of whether it has a radio identification.
It is clear that a careful, un-biased analysis of the
robustness of SCUBA detections in shallow maps is called for.  In this vein
we provide a careful estimate of flux boosting and have followed up in 
photometry mode 5 sources detected in a shallow $850\,\mathrm{\mu m}$ map
of the Groth Strip, in an attempt to quantify the amount of flux boosting present 
in the map.  To interpret the results we have developed a general method to 
assess the reliability of low SNR sources.

We apply these techniques to a particular SCUBA survey in the `Groth Strip'.  
The `Groth Strip Survey' (GSS) is a {\sl Hubble Space Telescope}
({\sl HST}) programme (GTO 5090, PI: Groth) consisting of 28
overlapping {\sl HST} Wide Field Planetary Camera 2 (WFPC2)
medium-deep images, covering an area of $113\,\mathrm{arcmin^{2}}$,
forming a long strip centred on
RA=\(14^{\mathrm{h}}16^{\mathrm{m}}38{\fs}8\),
Dec=\(52^{\circ}16'52''\) (J2000), at a Galactic latitude of
$b\simeq60^{\circ}$.  The GSS was the deepest {\sl HST}
cosmological integration before the HDF, reaching a limiting Vega
magnitude of $\sim27.5$--28 in both the \textit{V} and \textit{I}
bands \citep{Groth}.  The GSS has an enormous legacy value, since
extensive multi-wavelength observations centred on this region have
been conducted or are planned.  Morphological and photometric
information from the WFPC2 images are provided by the Medium Deep
Survey (MDS) database \citep{Ratnatunga} and the Deep Extragalactic
Evolutionary Probe (DEEP\footnote{See http://deep.ucolick.org}) survey
\citep{Simard}.  X-ray sources have also been identified in an
$80\,\mathrm{ks}$ {\sl XMM-Newton} observation of the GSS
\citep{Miyaji}.  The GSS is currently part of the on-going
DEEP2\footnote{See http://deep.berkeley.edu} survey and is also
targetted to be a major component of upcoming large surveys in the UV
(using the Galaxy Evolution Explorer, GALEX\footnote{See
http://www.galex.caltech.edu}), in the optical (as part of the
Canada-France-Hawaii Telescope Legacy Survey, CFHTLS\footnote{See
http://www.cfht.hawaii.edu/Science/CFHLS}), and in the IR (the
{\sl Spitzer} GTO IRAC Deep Survey).

In this paper, we present $850\,\mathrm{\mu m}$ SCUBA observations of
about 60 per cent of the original WFPC2 coverage of the GSS.  We have
also performed confirmation photometry on some of the sources.  Our
goal is to make the $850\,\mathrm{\mu m}$ map and source list
available to the community so that it may be correlated against
existing and future data sets at other wavelengths.
No claim is made that this survey
is either the deepest or the most extensive performed using SCUBA.
However, the observations cover enough integration time that we expect
a handful of real sources to be detected, and our survey represents
the best submillimetre data likely to be available in this field until
the advent of SCUBA-2.

\section{Map Observations and Data Reduction}\label{obs}
A roughly $70\,\mathrm{arcmin}^{2}$ portion of the Groth Strip (GSS)
was observed with a resolution of 14.7 arcsec and 7.5 arcsec at 850
and $450\,\mathrm{\mu m}$, respectively, with the 15-m JCMT atop Mauna
Kea in Hawaii in January 1999 and January 2000.  The GSS SCUBA map is
centred on RA$\,{=}\,14^{\mathrm{h}}16^{\mathrm{m}}00^{\mathrm{s}}$,
Dec$\,{=}\,52^{\circ}10^\prime00^{\prime\prime}$ (J2000).

52 overlapping 64-point jiggle maps of the GSS were obtained,
providing measurements of the continuum at both wavelengths
simultaneously with SCUBA \citep{Holland}, which has a field of view
of 2.3 arcmin.

The atmospheric zenith opacity at $225\,\mathrm{GHz}$ was monitored
with the Caltech Submillimetre Observatory (CSO) tau
($\tau_{\mathrm{CSO}}$) monitor.  The $\tau_{\mathrm{CSO}}$ ranged
from 0.03 to 0.09 in January 1999 and from 0.05 to 0.08 in January
2000.  The weather was generally more stable for the latter set of
data.

The secondary mirror was chopped at a standard frequency of $\simeq
8\,\mathrm{Hz}$ in azimuth to reduce the effect of rapid sky
variations.  The telescope was also `nodded' on and off the source.  A 40
arcsec chop--throw was used at a position angle of $54^{\circ}$,
almost parallel to the lengthwise orientation of the strip.  Pointing
checks were performed hourly on blazars and planets and varied by less
than 3 arcsec in azimuth and by less than 2 arcsec in elevation.
The overlapping jiggle maps were co-added to produce a final map with
a total integration time of 18 hours and 50 minutes.

We used SURF (SCUBA User Reduction
Facility; \citealt{Jenness}) scripts together with locally developed
code \citep{Borysthesis} to reduce the data.  The SURF map and our map look
similar.  The benefit of using our own code to analyse the
data is that it makes a map with minimally correlated pixels and
provides an estimate of the noise in each pixel.  We chose 3 arcsec
pixels oriented along RA,Dec coordinates.  This pixel size is slightly
too large for $450\,\mathrm{\mu m}$ studies, but has proven to be
adequate at $850\,\mathrm{\mu m}$ (see \citealt{Borys2003}).

\subsection{Flux Calibration}
Calibration data were reduced in the same way as the GSS data.  The
flux conversion factors (FCFs) over 3 of the 4 nights in January 1999
and all 3 nights in January 2000 agree with the monthly averages to
within 10 per cent (see the JCMT calibration web-page).  The
FCF value for one night in January was 30 per cent higher than the
monthly average and this could indicate that the sky was so variable
that the $\tau_{\rm CSO}$ was not accurately reflecting the opacity along
the line of sight to the object.  The calibration uncertainty is
omitted from our quoted error values since it is not a major
contributor to the global uncertainty of our low SNR data and has no
effect on our source detection method.

\begin{figure*}
\epsfig{file=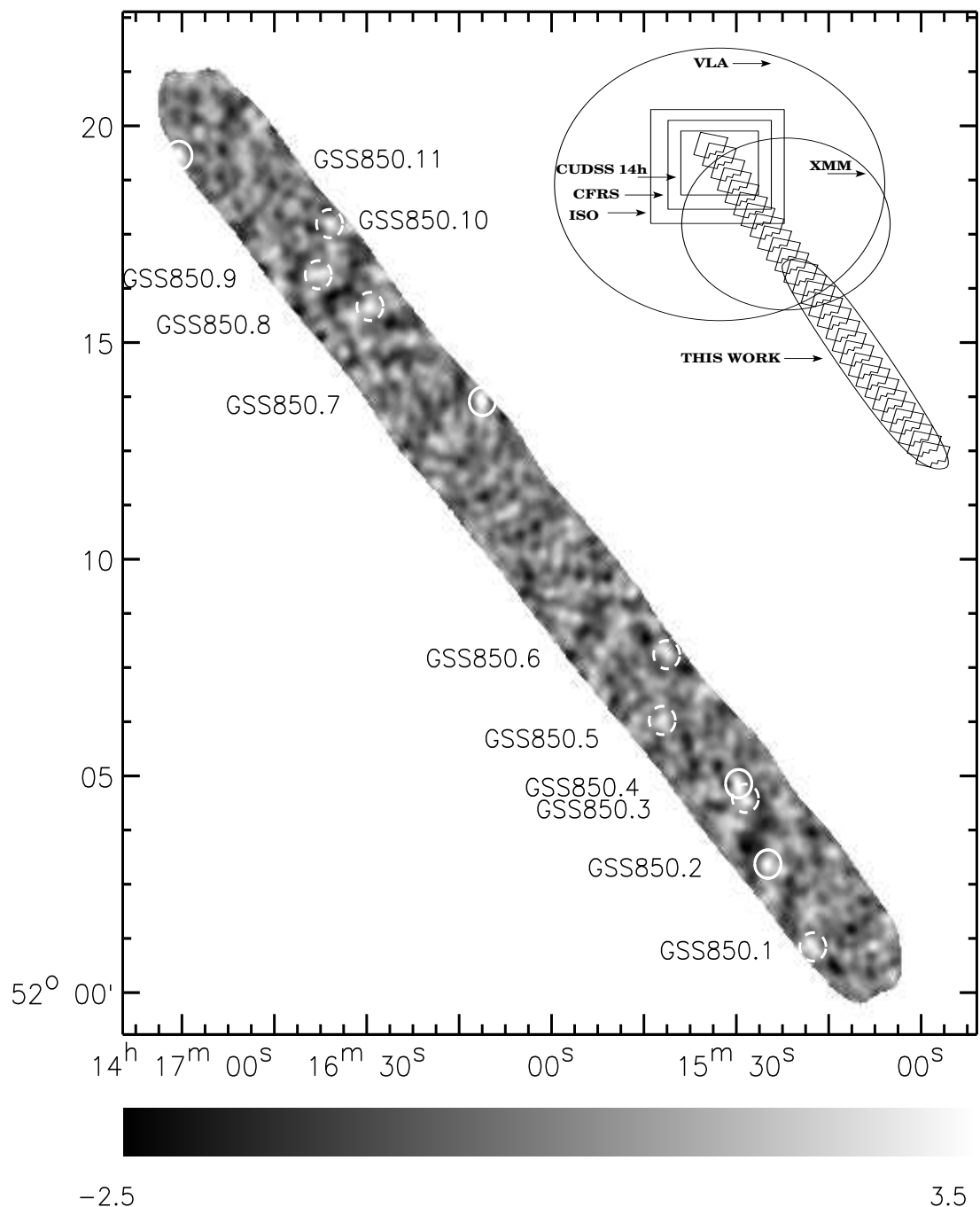,width=1.0\textwidth,scale=0.5}
\caption{The $850\,\mathrm{\mu m}$ SNR image of the Groth Strip,
smoothed with the 3-beam template (see $\S~\,\ref{sourcedet}$).  The
solid 40-arcsec diameter circles correspond to candidate sources with
$\mathrm{SNR}\geq3.5$.  Dashed circles indicate the
positions of the SNR$\,{=}\,3.0$--3.5 candidate sources.
The inlay in the top right-hand corner
illustrates our field geometry relative to some other surveys in this
region, including the original WFPC pointings (jagged squares), the CUDSS+14 field, 
CFRS and ISO survey regions (trio of large squares, listed here in order of increasing size), 
VLA coverage (large circle) and XMM coverage (smaller circle).}
\label{fig:850map}
\end{figure*}

The $850\,\mathrm{\mu m}$ map has a mean consistent with zero, as
expected from differential measurements, and an rms of
$3.5\,\mathrm{mJy}$.  The final map is shown in Fig.~\ref{fig:850map}.
The $450\,\mathrm{\mu m}$ map also has a mean consistent with zero and
an rms of $50\,\mathrm{mJy}$.

\subsection{Source Detection Method}\label{sourcedet}
Given the 14.7 arcsec beam, a high-redshift galaxy will be unresolved
and will appear as a positive source flanked by 2 negative sources.
The source density at $850\,\mathrm{\mu m}$ (see e.g. \citealt{Scott},
\citealt{Borys2003}, \citealt{Webb}) suggests that only a handful of sources
will be recovered in our map.  Hence we do not expect many overlapping sources,
and therefore sources were extracted by fitting the raw rebinned map
with a three-lobed PSF of an isolated point-source with the same chop throw and
position angle as the map data.  A fit to the PSF model centred on each pixel
is equivalent to a noise-weighted convolution and is the minimum variance estimator 
if the background consists of white noise.  An accompanying weighted noise map
is created simultaneously and provides an estimate of the noise
associated with the detection of a point source in each pixel.  A peak in the 
PSF-convolved map which is 3 times the noise in that pixel constitutes a 
$3\,\sigma$ detected point source.

\section{Source Robustness}

For a list of sources found above a given significance level in a map to be useful, 
one must address the following questions:  `Are there any statistical anomalies in the data which 
would cause one to doubt any of the sources?'; `Given our actual noise and measurement strategy, what 
fraction of sources present at any given flux level would we expect to detect?', that is, `How complete is our source list?'; and finally, `Do the fluxes inferred from the maps form a biased estimator of actual source flux?'  Bias and flux boosting are addressed in $\S~\ref{section4}$.  Quality of fit and completeness are addressed here.

We investigate source robustness using several techniques, including
spatial and temporal $\chi^{2}$ tests, searching for negative sources, and Monte Carlo simulations.

\subsection{Spatial and Temporal $\chi^{2}$ Tests}\label{chi}
Although a candidate source may be `detected' in the map, it may not
necessarily be well-fit by the PSF, or it may be a poor fit to the set
of difference data, or both.  We have performed spatial and temporal
$\chi^{2}$ tests in order to determine how well the raw timestream
data fit the final PSF-fitted maps.  See \citet{Pope} for details.

The spatial $\chi^{2}$ test provides a gauge across the map of the
goodness-of-fit of the triple-beam differential PSF to the data, and thus indicates if a
source is poorly fit by the assumed PSF.  We
find that the $\chi^{2}$ values for each pixel where a source is
detected are within $\pm2\,\sigma$ (=$2\sqrt{2N_{\mathrm{pix}}}$), where $N_{\mathrm{pix}}$ is the of number degrees 
of freedom or number of pixels included in the fit) in
all cases, except for GSS850.3, a candidate source with SNR=3.3.  The
poor fit may be the result of its proximity to GSS850.4 (see
Fig.~\ref{fig:850map}).

The temporal $\chi^{2}$ provides a measure of the self-consistency of
the raw timestream data which contribute flux to each map pixel.  For example, a portion of the hits on a pixel might be consistent with a certain flux value, whereas the rest of the hits might be most consistent with a different value.  Following the prescription of \citet{Pope}, we calculate the pixel temporal $\chi^{2}_i$ and the number of hits for each pixel $N_{\mathrm{hits}}^{i}$.  We then essentially construct a SNR map of poorness-of-fit to the model by using the quantity $\chi^{2}_i - N_{\mathrm{hits}}^{i}$ as the `signal' and $\sqrt{2N_{\mathrm{hits}}^{i}}$ as the `noise' and fitting the PSF to this temporal $\chi^{2}$ map.  We
find that none of the pixels at the centres of our candidate sources 
lie outside the $\pm 2\,\sigma$ regions of our distribution.  We are therefore confident
that all of our candidate sources lie in regions of the map with
self-consistent timestream data and have no grounds to reject any of
our detected candidate sources based on these tests.

Note that these two tests also check for Gaussianity of the noise in the map, 
but they are not a strong test of this distribution.

\subsection{Sources Detected in the Inverted Map}\label{invert}

A quick test of source reality is to create the negative of the
map and to search for sources using the same triple-beam template.
Aside from pixels associated with the off-beams of positive detections, we find 
6 `detections' in the inverted map, consistent with the expected number of false positive 
detections in noisy data (see $\S~\ref{sims}$).

\subsection{Monte Carlo Simulations}\label{sims}

Simulations are required in order to evaluate map completeness and the 
likely rate of false positive detections, given our non-uniform noise.  We follow the procedures
described in \citet{Borys2003} for investigating the number of
positive sources expected at random, as well as the map completeness.  
%We have also developed a new way of dealing
%with the effects of flux boosting, which we leave to $\S~\ref{boost}$.

\begin{figure}
\epsfig{file=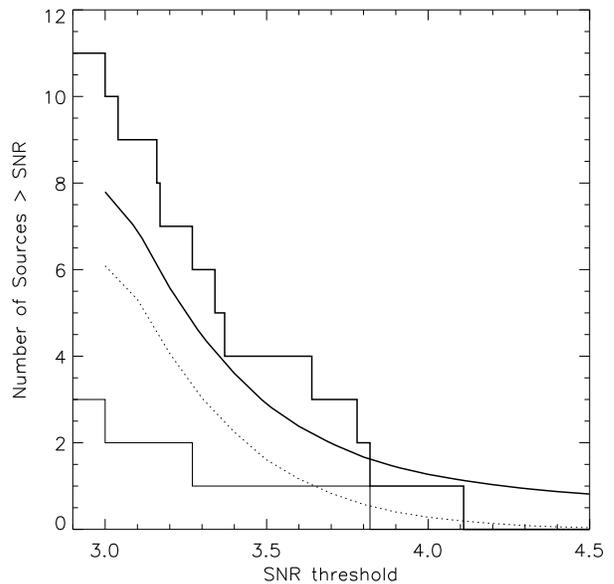,width=0.5\textwidth}
\caption{Cumulative number of detected candidate sources.  We plot
results against SNR threshold for our map (dark histogram), the
average expected in source-free simulated $850\,\mathrm{\mu m}$ maps
containing only Gaussian random noise (light dotted curve), the number
of false positives expected on average plus the predicted number of
real sources using the source counts of \citet{Borys2003} multiplied 
by the completeness estimate (dark solid curve), and the confirmed 
spurious detections (light histogram, see
$\S~\ref{section4}$).  The number of expected detections and the number of actual 
detections are consistent within the Poisson noise.}
\label{fig:fake}
\end{figure}

In order to determine how many detections may be spurious, we created
a map with the same shape and size as the real one, but replaced the
$850\,\mathrm{\mu m}$ data with Gaussian random noise, generated using
the rms of the timestream of each file and each bolometer.  We then
performed the same source detection procedure that was used on the
real data.  We repeated this sequence of steps 1000 times and plot the
cumulative number of positive sources detected on average in
Fig.~\ref{fig:fake} at each SNR threshold.  The simulations suggest
that on average one expects 1.6 false positive sources $>3.5\,\sigma$ and 
a further 4.5 between 3 and $3.5\,\sigma$.

\begin{figure}
\epsfig{file=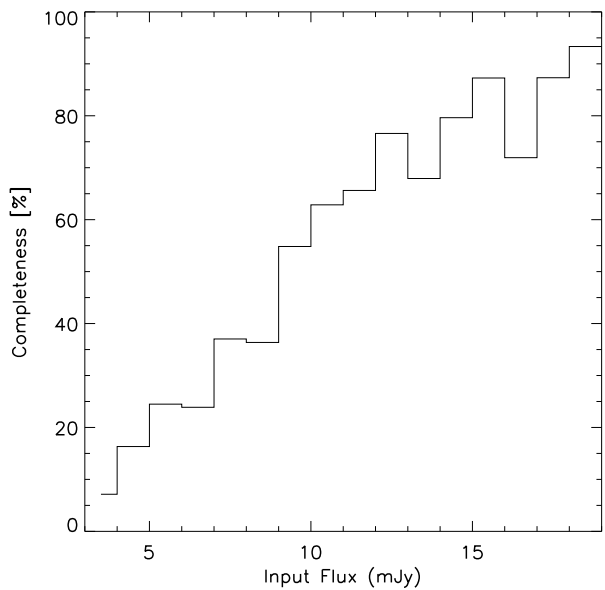,width=0.5\textwidth}
\caption{Completeness of $850\,\mu$m source recovery at each level 
of input flux for $3\,\sigma$ detections, as
determined from the Monte Carlo simulations of individual sources
added to the GSS data described in $\S~\,\ref{sims}$.}
\label{fig:sims}
\end{figure}

The completeness of a map is the fraction of sources which one expects to detect at each flux level.  
To measure this fraction, we
added a source of known flux into the real map and tried to extract it
using our source extraction method.  We selected the input source flux
randomly in the range 3--$20\,\mathrm{mJy}$, located it uniformly across the map, 
and repeated this procedure 1000 times.  A source is considered recovered if it is detected with
$\mathrm{SNR}\geq3$ and located within 7.5 arcsec (the
$850\,\mathrm{\mu m}$ beam HWHM) of the input position.  We estimate that
about 60 per cent of the $>10\,\mathrm{mJy}$ sources are detected
above a SNR of $3\,\sigma$ in the map (see Fig.~\ref{fig:sims}).  The completeness is 
slightly higher than it would be in a map with uniform noise at the same rms, as expected.

\section{Candidate Submillimetre Sources}\label{section4}

We detect 4 candidate sources with $\mathrm{SNR}\geq3.5\,\sigma$ and 7
candidate sources with SNRs in the range 3.0--$3.5\,\sigma$.  A
submillimetre image of the GSS is shown in Fig.~\ref{fig:850map}, 
where we number each of these 11 candidates, while
in Table~\ref{tab:sources} we present information on the 5 sources for
which we performed follow-up photometry (see $\S~\ref{phot}$).  The
signal and noise maps may be downloaded from
{\tt http://cmbr.physics.ubc.ca/groth}.

\begin{table}
\begin{minipage}{0.5\textwidth}
\caption{GSS $850\,\mathrm{\mu m}$ candidate submillimetre sources
with follow-up photometry.  Locations of these and the other 6 candidate SCUBA
sources are indicated in Fig~\ref{fig:850map}.\label{tab:sources}}
\begin{tabular}{lcc}
\hline \multicolumn{1}{c}{Object} &
\multicolumn{1}{c}{$S_{850}$\footnote{Flux density estimate from the
map.}} & \multicolumn{1}{c}{$S_{850}$\footnote{Flux density measured
from follow-up photometry.}}\\ \multicolumn{1}{c}{} &
\multicolumn{1}{c}{(mJy)} & \multicolumn{1}{c}{(mJy)}\\\hline GSS850.1
& $9.5\pm2.9\,(3.3\,\sigma)$ & $-1.8\pm3.6\,(-0.5\,\sigma)$\\ GSS850.2
& $8.2\pm2.3\,(3.6\,\sigma)$ & $5.7\pm1.1\,(5.2\,\sigma)$\\ GSS850.6 &
$12.3\pm4.1\,(3.0\,\sigma)$ & $1.2\pm3.1\,(0.4\,\sigma)$\\ GSS850.7 &
$13.2\pm3.2\,(4.1\,\sigma)$ & $4.2\pm1.7\,(2.5\,\sigma)$\\ GSS850.11 &
$11.2\pm2.9\,(3.8\,\sigma)$ & $0.1\pm1.6\,(0.1\,\sigma)$\\
\end{tabular}
\end{minipage}
\end{table}

\subsection{Additional Photometry}\label{phot}

In December 2003 and January 2004, SCUBA photometry observations were
performed in the 2-bolometer chopping mode to check some of our candidate
map detections.  We selected 2 candidates, GSS850.7 and GSS850.11,
near the noisier edge regions of the map and a `control', GSS850.2, in
a lower-noise region away from the edges.  Also, during one of the
observing runs in January 2000, four sources were identified
in the map data `by eye' and selected as targets for follow-up
photometry.  Only two of these pointings correspond to candidate detections in
the final map (GSS850.1 and GSS850.6); this is a warning that our eyes
often pick out bright outliers in noisy regions of a map.  All of the
photometry observations were reduced in the standard way using SURF.
In order to increase the SNR of sources observed in the 2-bolometer
mode by a factor of approximately $\sqrt{3/2}$, we folded in the
signal from the off-position bolometers to the central bolometer (see
\citealt{Chapman2000}).  The results are listed in
Table~\ref{tab:sources} for comparison with the estimated map fluxes.
In all cases, the photometry pointings were within 3 arcsec of the positions found 
by the source-detection algorithm in the
map.

\subsection{Flux Boosting in the Map}\label{boost}

It appears that we have detected some sources in the map.  However
none of the candidate sources have very high SNRs, and so we need to
be careful in interpreting these results.  Confusion can
either increase or decrease the input flux of a source, but a noisy
flux-limited map will preferentially contain sources whose true fluxes have
been increased (usually called Malmquist bias) and this effect is
exacerbated for steep source counts.  Our source extraction procedure
therefore biases fluxes in the map upwards, which we now
attempt to quantify.

\begin{figure*}
\epsfig{file=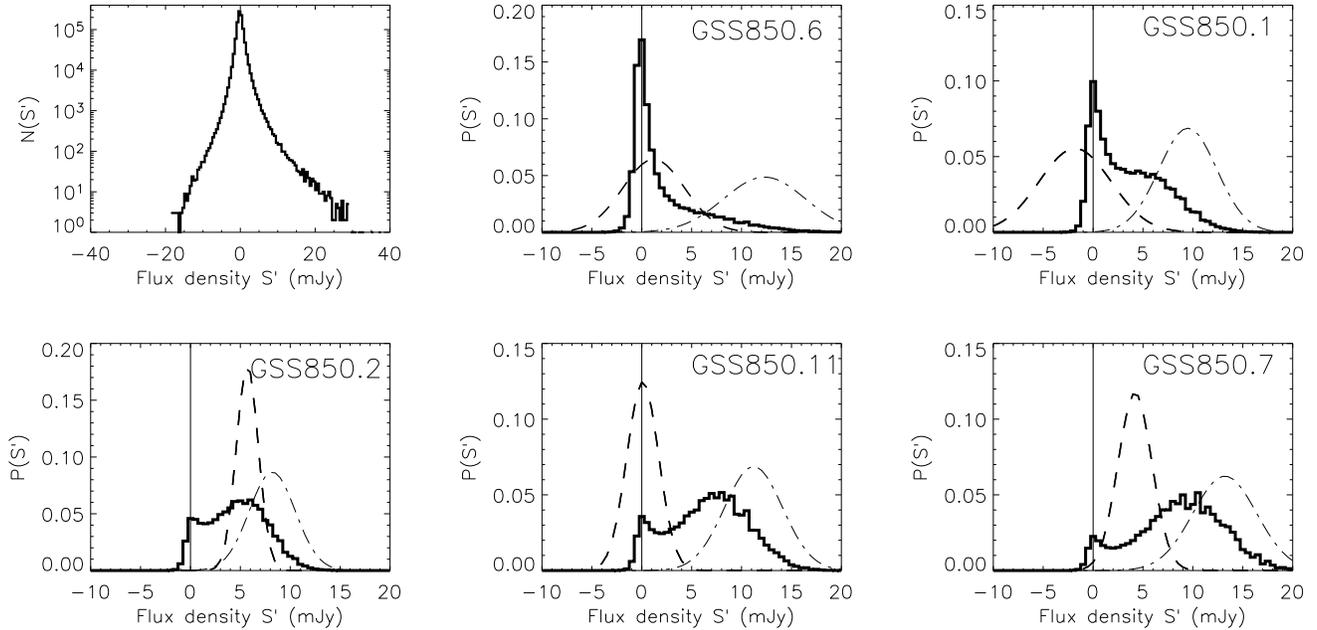}
\caption{The first panel shows the histogram of the 1
million noiseless triple-beam simulations described in $\S~\ref{boost}$.  
This is the $P(D)$ distribution (e.g. \citealt{Scheuer}, \citealt{Condon}) 
for a triple-beam experiment, which is strongly peaked around zero and skew 
positive due to sources.  The other panels show the resulting flux probability distributions
(dark histograms) for the 5 map source candidates which had follow-up
photometry observations, given their measured fluxes and errors
(assumed to be Gaussian distributions, plotted separately as light
dot-dashed lines) and the underlying source count model.  Photometry 
measurements are overplotted for comparison and are also assumed
to have Gaussian probability distributions (dark dashed lines).  A
vertical line at $S_{\mathrm{p}}$=0 is plotted as a reference with which to compare the
photometry measurements.  One can see that the
combination of the intrinsic distribution of fluxes and the low SNR
measurements from the map make the photometry measurements much less
inconsistent than they might appear (i.e.~the dashed Gaussians compare well
with the solid histograms, even although the two Gaussians in each panel are
usually discrepant).}
\label{fig:boost}
\end{figure*}

We performed a set of simulations to assess the expected distribution
of pixel brightnesses from triple-beam (i.e.~double difference)
observations of a noiseless blank-sky.  We used a smooth curve fit to 
(and mildly extrapolated from) the number counts of
\citet{Borys2003} as an a priori distribution of fluxes in the range
0.1--$40\,\mathrm{mJy}$.  We then populated 3 different patches of noiseless
sky following a Poisson distribution, and sampled each area with a
SCUBA Gaussian beam, taking a double difference each time.  This was
done 1 million times and the anticipated prior distribution of double
difference flux measurements, $N(S_{\mathrm{p}})$, is plotted in the first panel of
Fig.~\ref{fig:boost}.  We have also investigated the effects of reasonable excursions 
from the assumed shape of the source counts.  For example, using the $1\,\sigma$ error 
bar values at the bright end of the number counts has less than a 10 per cent 
effect on the resulting flux estimates.

The distribution $N(S_{\mathrm{p}})$, which is the prior probability that a pixel in the map has differential flux $S_{\mathrm{p}}$, is calculated from \textit{noiseless} simulations while our actual map contains noise.  If $M$ is the statement that we measure flux $S_{\mathrm{m}}\pm\,\sigma_{\mathrm{m}}$ at some pixel in the map, the probability that the true flux of that pixel is $S_{\mathrm{p}}$ is obtained from Bayes' theorem:

\begin{equation}
P(S_{\mathrm{p}})\equiv P(S_{\mathrm{p}}|M,N(S_{\mathrm{p}}))=\frac{N(S_{\mathrm{p}})\times P(M|S_{\mathrm{p}})}{P(M)},
\label{eqn:pd}
\end{equation}
where the probability we would measure $S_{\mathrm{m}}$ when the true flux is  $S_{\mathrm{p}}$ is
\begin{equation}
P(M|S_{\mathrm{p}})=A\,e^{\frac{(S_{\mathrm{m}}-S_{\mathrm{p}})^{2}}{2\,\sigma_{\mathrm{m}}^{2}}},
\end{equation}
under the assumption that our noise is Gaussian distributed.  We have weakly tested this assumption in $\S~\ref{chi}$ $\mathrm{and} ~\ref{invert}$.  $P(M)$ acts as an overall normalisation in equation (\ref{eqn:pd}), and does not depend upon $S_{\mathrm{p}}$ as long as the noise is not correlated with sources on the sky.  Strictly speaking, $P(M|S_{\mathrm{p}})$ should be altered from the form we have used to account for the fact that we are examining the probability at a location where we have found a peak.  In practice, at $S_{\mathrm{m}}-S_{\mathrm{p}} \geq 3\,\sigma$ the full expression converges to the simpler form we have used (\citealt{Bond}).

The posterior flux probability distribution, $P(S_{\mathrm{p}})$ is shown as a solid histogram in the panels of Fig.~\ref{fig:boost} for each of the 5 sources for which we also have follow-up photometry information.  The dot-dashed Gaussian in each panel is $P(M|S_{\mathrm{p}})$, which is often incorrectly adopted as the flux estimate of a map source.

In Fig.~\ref{fig:boost}
we have placed the individual $P(S_{\mathrm{p}})$ plots in 
order of increasing map SNR.  It is clear that one expects to
measure a non-zero flux value a significant fraction of the time only
for sources with relatively high SNRs.  Sources with modest SNRs are
much more likely to have non-zero photometry results than lower SNR
sources.  The peak in the a posteriori distribution at zero flux dominates for SNR $\lesssim3.5$.  This confirms the usual prejudice towards high SNR sources -- if a source is bright, it needs to be detected with a
$\mathrm{SNR}\gtrsim4\,\sigma$ in order to be deemed a secure
detection.  Moreover, we also find that at the same SNR level,
apparently brighter sources (with consequently higher flux
uncertainty) are more likely to be spurious (i.e. flux boosted from
$\simeq0\,\mathrm{mJy}$) than fainter sources.  Thus at a given SNR, low flux 
sources are more likely to be real than high flux sources.  

We performed independent photometry observations on 5 sources with SNRs
ranging from 3.0--$4.1\,\sigma$ and the results are shown in 
Table~\ref{tab:sources}.  We wish to know if the photometry
measurements are consistent with the map-detected fluxes.  In other
words, we want to answer the question: `What is the probability that
we will measure $S_{\mathrm{p}}$ in photometry mode given the
map-detected flux ($S_{\mathrm{m}}$) and uncertainty
($\sigma_{\mathrm{m}}$) and the underlying source count model?'.  
The main point is that the a posteriori probability of finding a 
bright source will be down-weighted by the a priori probability coming from the source counts.

A comparison of the dashed curves and solid histograms in Fig.~\ref{fig:boost} shows that our photometry results are consistent with $P(S_{\mathrm{p}})$, even though the photometry is often inconsistent with the raw (i.e.~flux boosted) map readings $P(M|S_{\mathrm{p}})$.

\subsection{A Revised Source List}

We can now assess the probability of obtaining each of the photometry
measurements using the distributions in Fig.~\ref{fig:boost}.  For
GSS850.6, a photometry result \textit{lower} than what we measured is
expected 43 per cent of the time.  For the remaining sources GSS850.1,
GSS850.2, GSS850.11 and GSS850.7, we have assessed that a photometry
measurement lower than the one we obtained would have occurred 12, 60,
6, and 18 per cent of the time, respectively.  A  Kolmogorov-Smirnov (KS) 
test performed on these results determined that the set of 5 trials is 
consistent with a uniform distribution.

The photometry results are thus completely
within the realm of what is expected, despite the apparently contradictory 
results presented in Table~\ref{tab:sources}.  We confirm GSS850.2 and 
GSS850.7 as bona fide $850\,\mathrm{\mu m}$
sources, and we confidently eliminate GSS850.1, GSS850.6, and
GSS850.11 from our candidate source list.  Note that these eliminated
sources are also among the 4 noisiest candidate detections in the map, which
makes it even less surprising that they are spurious.  We have
tallied up the number of detections, expected sources, and spurious
detections and have illustrated these in Fig.~\ref{fig:fake}.  We note
that the number counts (e.g. \citealt{Scott}, \citealt{Borys2003},
\citealt{Webb}) predict the detection of around 3 sources at the flux
limit of the map ($\sim10\,\mathrm{mJy}$).

\begin{table*}
\begin{minipage}{\textwidth}
\caption{GSS $850\,\mathrm{\mu m}$ revised source list.  We
have included the 2 photometry-confirmed sources and an additional
$>3.5$ candidate source which has a reasonable chance of being real.
The best $850\,\mathrm{\mu m}$ flux estimate is given based on the 
combination of the posterior probability of the map flux given the data together with
 the photometry flux for the sources with photometry (2 and 7), and based solely 
on the posterior probability of the map flux for GSS850.4.  The reported flux is the most likely flux in the 68 per cent confidence region, with upper and lower error bars shown to indicate the range of that confidence interval.  95 per cent Bayesian upper limits are also given for the
$450\,\mathrm{\mu m}$ flux of each $850\,\mathrm{\mu m}$
detection.\label{tab:confirmedsources}}
\begin{tabular}{lcccl}
\hline \multicolumn{1}{c}{Object} & \multicolumn{2}{c}{Position
(2000.0)} & \multicolumn{1}{c}{$S_{850}$} &
\multicolumn{1}{c}{$S_{450}$}\\ \multicolumn{1}{c}{} &
\multicolumn{1}{c}{RA} & \multicolumn{1}{c}{Dec} &
\multicolumn{1}{c}{(mJy)} & \multicolumn{1}{c}{(mJy)}\\\hline 
GSS850.2 & \(14^{\mathrm{h}}15^{\mathrm{m}}25{\mathrm{\fs}}0\) & \(+52^{\circ}02'57''\) & $5.9^{+0.7}_{-1.2}$ & $<170$\\ 
GSS850.4 & \(14^{\mathrm{h}}15^{\mathrm{m}}29{\mathrm{\fs}}5\) & \(+52^{\circ}04'48''\) & $6.9^{+2.0}_{-4.7}$ & $<48$\\ 
GSS850.7 & \(14^{\mathrm{h}}16^{\mathrm{m}}11{\mathrm{\fs}}6\) & \(+52^{\circ}13'42''\) & $4.8^{+1.5}_{-1.9}$ & $<118$ \\
\end{tabular}
\end{minipage}
\end{table*}

Our revised source list now includes 2 confirmed sources (with
coordinates given in Table~\ref{tab:confirmedsources}) as well as 1
other candidate which is $>3.5\,\sigma$ in the map.  Although not
confirmed by photometry, Figs.~\ref{fig:fake} and~\ref{fig:boost} suggest that SNR $>3.5$ sources 
have a reasonable chance of being real.  We are able to calculate 
a best estimate of the flux for each of these objects using the 
combination of all available information, including the map measurements, 
photometry measurements and the source count prior.  To do this we multiply the 
measured (assumed Gaussian) photometry flux probability distribution, 
$P(S_{\mathrm{p}},\sigma_{\mathrm{p}})$, 
by the calculated posterior probability for the map flux (equation (\ref{eqn:pd})), which we take as the prior distribution for $S_{\mathrm{p}}$, and we normalise it to have unit integral.  For the candidate 
source, GSS850.4, we only know the a posteriori distribution for the map measurement, since we do not have a photometry measurement for this object.  In Table~\ref{tab:confirmedsources} we give the peak of these new distributions, along with the error bars describing the 68 per cent confidence regions.

\section{A First Attempt at Multi-wavelength Correlations}

We now use our new candidate source list (see
Table~\ref{tab:confirmedsources}) to search for close counterparts at
other wavelengths in other data sets which overlap with our coverage.
We also perform stacking analyses to see if there is any overlap
between the catalogues and maps.

The $450\,\mathrm{\mu m}$ map of this region is of poor quality; the data
are shallow (since the sensitivity at $450\,\mathrm{\mu m}$ is worse)
and inhomogeneous (being more prone to changes in the weather).
We do not detect any of our $850\,\mu$m sources in the $450\,\mathrm{\mu m}$
map, but we present 95 per cent confidence upper limits to the
$450\,\mathrm{\mu m}$ flux for each $850\,\mathrm{\mu m}$ detection in
Table~\ref{tab:confirmedsources}.  The $450\,\mathrm{\mu m}$ average (or `stacked') flux
density at the 3 $850\,\mathrm{\mu m}$-detected positions is $10\,(\pm23)\,\mathrm{mJy}$.

Using an $80\,\mathrm{ksec}$ {\sl XMM-Newton} observation
encompassing the northeast part of the GSS, \citet{Miyaji} have
uncovered about 150 sources down to flux limits of $\simeq 1\times
10^{-20}$ and $\simeq 2\times 10^{-20}\,\mathrm{W}\,\mathrm{m^{-2}}$
in the soft (0.5--$2\,\mathrm{keV}$) and hard (2--$10\,\mathrm{keV}$)
X-ray bands, respectively.  Of these detections, 7 lie within our
submillimetre map and the X-ray positional errors are typically about
2--3 arcsec.  No X-ray counterparts exist within the anticipated error
circle of 4", and indeed even there are no counterparts within a full
beam of any SCUBA source.  However, the
stacked $850\,\mathrm{\mu m}$ flux from the 7 X-ray positions lying
within the submillimetre map region is $2.5\,(\pm1.1)\,\mathrm{mJy}$.
This corresponds to a $0.8\,\mathrm{mJy}$ 95 per cent confidence \textit{lower} limit to the
mean flux of these sources.  

These X-ray sources are therefore brighter than Lyman-break galaxies at  $850\,\mathrm{\mu m}$ (e.g.~\citealt{Chapman2000})!  If AGNs do not comprise a large fraction of our sources, this 
result indicates that the X-ray emission originates from processes related to star-formation.  
This result illustrates that this map can, in fact, be used to make statistical remarks about
$\sim1\,\mathrm{mJy}$ sources even though individual detections are
hopeless, and shows a path to populating the confusion sea in the
submillimetre (see also \citealt{Borys2004}).

\section{Conclusions}\label{conc}

We have mapped approximately $70\,\mathrm{arcmin^{2}}$ of the Groth
Strip at $850\,\mathrm{\mu m}$ with SCUBA on the JCMT to a $1\,\sigma$ 
depth of around $3.5\,\mathrm{mJy}$.

Using a robust source detection algorithm, we have found 11 candidate
sources with $\mathrm{SNR}\geq3\,\sigma$.  Monte Carlo simulations
suggest that most of these will either be spurious or considerably
flux boosted.  Follow-up photometry observations have confirmed 2 of
them and rejected 3.  Based on these follow-up photometry data, we
have determined, not surprisingly, that candidate sources in
high-noise regions of the map have implausibly high apparent fluxes at
SNR $\geq3\,\sigma$, and are likely to be spurious false-positive
detections.  We reiterate that bright sources detected in a map should
have $\mathrm{SNR}>3.5\,\sigma$ before they have a reasonable chance
of being real, and $\mathrm{SNR}>4\,\sigma$ before they should be
believed with any confidence.  Our final source list for the GSS
contains 2 confirmed SCUBA sources and 1 further candidate source with
$\mathrm{SNR}>3.5\,\sigma$.  Using a combination of the unboosted map flux posterior 
probability distributions and the photometry measurements (when available), we present 
best estimates of the flux for these objects.

We have measured a mild statistical detection of low flux
(${\sim}\,1\,\mathrm{mJy}$) sources at X-ray wavelengths through a
stacking analysis, and it may be that similar comparisons with data at
other wavebands might also be fruitful.  We have therefore made our
maps available at {\tt http://cmbr.physics.ubc.ca/groth}.

Our simple Bayesian method for correcting the effects of flux boosting should be
useful for future surveys such as SHADES and those carried out with
SCUBA-2, as well as for other instruments which provide data in the
low SNR near-confusion regime.  It may also be useful to adapt this method 
in order to find sources, by searching for pixels in a map for which the posterior probability 
for $S_{p}>0$ is above some threshold.

\section{Acknowledgments}\label{ack}
This work was supported by the Natural Sciences and Engineering
Research Council of Canada.  We would like to thank the staff of the
JCMT for their assistance with the SCUBA observations.  KC would like
to thank Vicki Barnard for assistance determining pre-upgrade
calibration FCFs and Bernd Weferling for assistance in determining
problematic $\tau_{\mathrm{CSO}}$ fits.  We also wish to thank an anonymous 
referee for constructive comments.  The James Clerk Maxwell Telescope is operated on behalf of
the Particle Physics and Astronomy Research Council of the United
Kingdom, the Netherlands Organisation for Scientific Research, and the
National Research Council of Canada.

\end{document}